\documentclass[letterpaper, 10 pt, conference]{ieeeconf}  

\IEEEoverridecommandlockouts                              
\usepackage{color}

\definecolor{orange}{rgb}{1.0,0.3,0.0}
\definecolor{violet}{rgb}{0.75,0,1}
\definecolor{darkgreen}{rgb}{0,0.6,0}
\definecolor{cyan}{rgb}{0.2,0.7,0.7}
\definecolor{blueish}{rgb}{0.2,0.2,0.8}
\definecolor{darkblue}{rgb}{0.1,0.1,0.9}
\definecolor{lightgray}{gray}{0.9}
\definecolor{hanpurple}{rgb}{0.32,0.09,0.98}

\usepackage{graphicx} 
\usepackage{flushend} 
\usepackage[colorlinks=true,linkcolor=black,anchorcolor=black,citecolor=black,filecolor=black,menucolor=black,runcolor=black,urlcolor=black]{hyperref}

\title{
\textsf{BOSS-LDG}: A Novel Computational Framework that Brings Together \\ \textsf{B}lue Waters, \textsf{O}pen \textsf{S}cience Grid, \textsf{S}hifter and the \textsf{L}IGO \textsf{D}ata \textsf{G}rid \\ to Accelerate Gravitational Wave Discovery}

\author{E.~A.~Huerta$^{1}$, Roland Haas$^{1}$,
Edgar Fajardo$^{2}$, Daniel S. Katz$^{1}$, \\ Stuart Anderson$^{3}$, Peter Couvares$^{3}$, Josh Willis$^{4}$, Timothy Bouvet$^{1}$ \\ Jeremy Enos$^{1}$, William T. C. Kramer$^{1}$, Hon Wai Leong$^{1}$ and David Wheeler$^{1}$\\ \\
$^{1}$NCSA, University of Illinois at Urbana-Champaign, Urbana, Illinois 61801, USA \\
\{elihu, rhaas, dskatz, tbouvet, jenos, wtkramer, hwleong, dwheeler\}@illinois.edu\\
$^{2}$University of California, San Diego, La Jolla, California 92093, USA \\
emfajard@ucsd.edu\\
$^{3}$LIGO, California Institute of Technology, Pasadena, California 91125, USA \\
\{anderson, peter.couvares\}@ligo.caltech.edu\\
$^{4}$Abilene Christian University, Abilene, Texas 79699, USA\\
josh.willis@acu.edu
}

\begin{document}

\maketitle
\thispagestyle{empty}
\pagestyle{empty}

\begin{abstract}
We present a novel computational framework that 
connects Blue Waters, the NSF-supported, leadership-class
supercomputer operated by NCSA, to the Laser Interferometer Gravitational-Wave
Observatory (LIGO) Data Grid 
via Open Science Grid technology. To enable this computational infrastructure, we 
configured, for the first time, a \textit{LIGO Data Grid Tier-1 Center} that can submit heterogeneous LIGO workflows using Open 
Science Grid facilities. In order to enable a seamless connection between the 
LIGO Data Grid and Blue Waters via Open Science Grid, we utilize \textit{Shifter} to containerize LIGO's workflow software. This work represents the first time Open Science Grid, Shifter, and Blue Waters are unified to tackle a scientific problem and, in particular, it is the first time a framework of this nature is used in the context of large scale
gravitational wave data analysis. This new framework has been used in the last several weeks
of LIGO's second discovery campaign to run the most computationally demanding gravitational 
wave search workflows on Blue Waters, and accelerate 
discovery in the emergent field of gravitational wave 
astrophysics. We discuss the implications of this novel framework for a wider ecosystem of Higher Performance Computing users.
\end{abstract}

\section{Introduction}

Some of the most extraordinary events in the Universe 
are driven by strong gravitational interactions. 
Mergers of black holes, 
neutron stars and white dwarfs can be used as
astrophysical laboratories to gain 
insights into the physics of objects moving at 
relativistic speeds in the presence of extreme
gravitational fields, the arena of Einstein's 
theory of general relativity~\cite{gr2}. 

General relativity predicts that mergers of ultra compact
objects produce a type of radiation that consists
of curvature fluctuations\footnote{Curvature is 
the manifestation of gravity in Einstein's theory
of general relativity.}, which travel unimpeded in
spacetime at the speed of light~\cite{enrosen,gr}. 
This radiation, known as gravitational waves, 
removes energy and angular momentum from compact binary
sources, driving the components of a system of two
orbiting objects  into an inspiral 
trajectory and eventually to collision and merger. Given the
complexity of 
Einstein's equations, a detailed study of this prediction 
led to the creation of a new field of
research---numerical relativity---which strongly relies 
on advanced High Performance Computing (HPC) facilities
to numerical study the physics of black holes, neutron stars, and other promising sources of gravitational waves~\cite{preto}.

Over the last decade, numerical relativity software
has steadily evolved and attained sufficient maturity to
generate catalogs of compact object mergers with
relative ease, which has enabled detailed studies of 
the physics of gravitational wave sources~\cite{Mroue:2013,2016CQGra..33t4001J,2017arXiv170303423H,Huerta:2017a}.
State-of-the-art cyberinfrastructure facilities, such as the 
Extreme Science and 
Engineering Discovery Environment (XSEDE) and the Blue 
Waters supercomputer, have been instrumental to realize this work, and push the frontiers of theoretical and computational astrophysics~\cite{2017arXiv170300924J}. 
 
In preparation for the detection of gravitational waves, the numerical relativity community shared catalogs of numerical relativity waveforms, which describe black hole mergers, with members of the Laser Interferometer Gravitational wave Observatory (LIGO) Scientific Collaboration (LSC). LIGO scientists used these numerical relativity waveforms to carefully assess whether their gravitational wave detection algorithms were capable of extracting them from highly noisy LIGO data~\cite{2012CQGra..29l4001A,2009CQGra..26p5008A,2014CQGra..31k5004A}. These studies were crucial to further develop and perfect the LSC Algorithm Library Suite (LAL)---a suite of various gravitational wave data analysis routines written in \texttt{C}~\cite{LAL}. 
 
In parallel to these developments, experimental physicists turned LIGO into the world's largest and most sensitive interferometric gravitational wave detector~\cite{LSC:2015}. After a 5-year US\$200-million upgrade, the advanced LIGO (aLIGO) detectors started collecting data in mid-September 2015~\cite{DII:2016}.

Another critical element for the success of the aLIGO mission, and the focus of this article, is the exploitation of state-of-the-art computational resources to enable the detection of gravitational waves. aLIGO data analysis is a compute-intensive science, which consumes hundreds of millions of CPU core-hours per year. Most of these analyses are embarrassingly parallel High Throughput Computing (HTC) work. Traditionally, LIGO data analysis computing has been done on the LIGO Data Grid (LDG), which consists of dedicated HTC clusters at seven LSC sites in the US and Europe, including the LIGO Laboratory\footnote{An LDG Tier-3 Center is a small computing cluster that can only be used by a limited set of LIGO researchers. An LDG Tier-1 Center is a large computing cluster that all members of the LSC can use.}. 

The LIGO mission is a prime example of a transdisciplinary research program that brought together complementary fields of research---theoretical and computational astrophysics, experimental physics, HPC and HTC---to enable the discovery of gravitational waves, and to establish an entirely new field of research that is revolutionizing the landscape of theoretical and observational astrophysics~\cite{DI:2016,secondec,thirddetection}. 

Anticipating that the lifetime of gravitational wave discovery campaigns will be longer, therefore producing larger datasets, and that gravitational wave detectors will be gathering data in four different continents within the next few years, LIGO is adopting recent developments
in HPC to leverage advanced cyberinfrastructure
facilities such as XSEDE and 
Open Science Grid (OSG) to accelerate gravitational wave discovery~\cite{2017Weitzel}.

\begin{figure}[tb]
   \centering
   \includegraphics[width=0.75\linewidth]{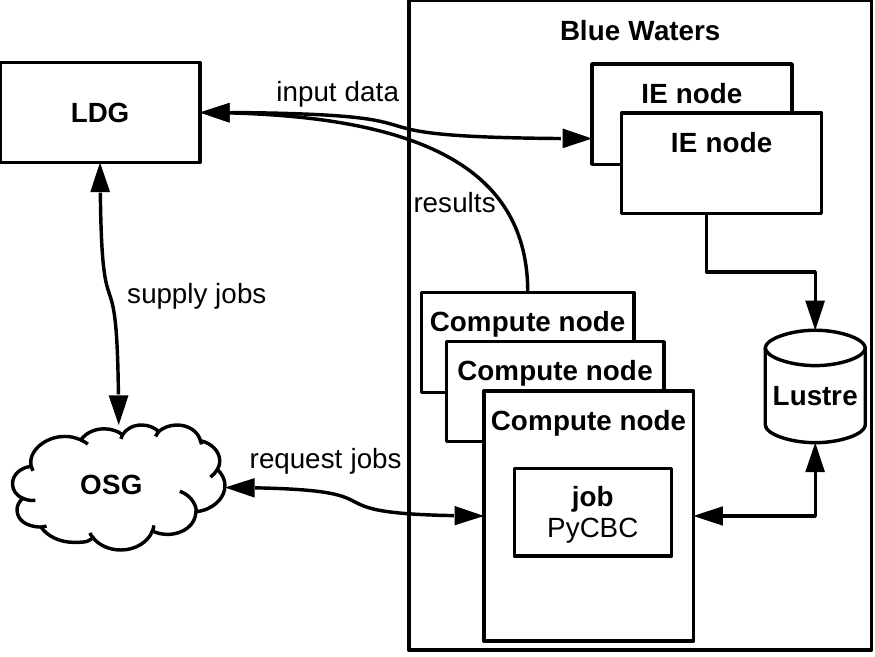}
  \caption{Interaction between the LIGO Data Grid (LSG), Open Science Grid (OSG) and Blue Waters. Import/Export (IE) nodes are
  Blue Waters' dedicated nodes that are used for file transfer in the shared file system. 
   \label{fig:osg_plot}}
\end{figure}

In this article, we present a novel computational framework (see Figure~\ref{fig:osg_plot}) that brings together joint efforts by NCSA, San Diego Supercomputer Center and LIGO scientists to connect the LDG to Blue Waters via OSG. The Blue Waters supercomputer is ideally suited to facilitate large-scale gravitational wave data analysis because the large number of independent jobs in these analyses can quickly be run on Blue Waters using the reasonably-large set of otherwise unoccupied nodes (through backfill). This
work has been accomplished in several steps:
(i) configuration of the first
LDG Tier-1 Center to run heterogeneous gravitational wave search
workflows via OSG;
(ii) 
implementation of \texttt{Shifter} to configure Blue Waters
as an OSG resource, and deal for the first time with
cybersecurity constraints, such as two factor authentication,
to allow incoming LDG jobs from Blue Waters authorized users to run on Blue Waters;
(iii) optimization of containers to significantly improve
the effectiveness of Blue Waters computing resources  for LDG type jobs
to accelerate time to discovery. \textit{This work represents the first time OSG, Shifter, and Blue Waters are unified to tackle a scientific problem and, in particular, it is the first time a framework of this nature is used in the context of large scale
gravitational wave data analysis.}

This article is organized as follows: Section~\ref{work} describes the construction
of the first framework that uses Shifter to enable the Blue Waters supercomputer as an 
OSG resource, and its specific 
application for gravitational wave data analysis. 
In Section~\ref{check}, we show that results obtained with  gravitational wave workflows using Blue Waters are consistent with results obtained  on traditional 
OSG resources. In Section~\ref{scale}, we discuss computational challenges that we encountered and overcame when running LIGO
workflows on Blue Waters. Section~\ref{application} presents 
the first use of \texttt{BOSS-LDG} to search for gravitational wave transients in aLIGO's second discovery campaign.
Section~\ref{related} lists other projects that use HPC resources to run similar
workflows and discusses differences of this approach. In Section~\ref{end} we describe the compatibility requirements that scientific workflows need to meet to benefit from this new framework, a summary of our work, and
future directions of research and applications of this
computational framework. 

\section{The LIGO Data Grid meets Blue Waters}
\label{work}

The computational framework we present in this article provides three significant benefits, one to LIGO, one to Blue Waters, and one to the overall cyberinfrastructure community.
 
First, it provides LIGO with significant computational resources to scale up its processing, and to promptly validate future potential major scientific discoveries. The existing LDG has sufficient resources to keep up with its current regular flow of work, however future gravitational wave discovery campaigns will be longer and will involve more detectors. Thus, additional computational resources will be needed to accelerate large scale gravitational wave searches. Furthermore, to support urgent needs, such as the detection of gravitational waves that are accompanied by emission of light and neutrinos, LIGO will need to rely on resources beyond those available for its normal processing to validate these discoveries. While keeping up with the regular flow of work may be sufficient for core investigations, \textit{ad hoc} and additional investigations based on core results may lead to new and detailed understanding that would otherwise will remain inaccessible. This work allows LIGO to use Blue Waters for these three purposes.
 
Second, computationally demanding workflows allow Blue Waters to increase cluster utilization and throughput by enabling tasks to backfill unused Blue Waters nodes. Because the LIGO tasks are independent from each other (they are not network sensitive or part of a tightly-coupled workload), Blue Waters can use the COMMTRANSPARENT flag when scheduling them, so that each task can be placed anywhere within the torus network without affecting the network performance of other jobs, and increasing the overall system utilization. We are adding HTC jobs to Blue Waters without decreasing the number of HPC jobs that the system can run.
 
Third, this work demonstrates the interoperability of NSF cyberinfrastructure resources, and shows how large projects can benefit from making use of existing resources rather than having to build their own custom solutions for all possible needs.

In the following subsections, we present a brief description of the Blue Waters supercomputer; the Open Science Grid; LIGO's computing needs over the next few years; one specific LIGO workflow; how the construction of the computational framework presented in this article adds, for the first time, large scale gravitational wave data analysis to the Blue Waters supercomputer scientific portfolio.

\subsection{Blue Waters}
\label{bluewaters}

Blue Waters is one of the largest supercomputers accessible to
academic researchers worldwide. 
It has 22,636 XE compute nodes, each containing two CPUs that use an x86 instruction set architecture (ISA) and 4,228 XK compute nodes, each of which contains one CPU that uses
an x86 ISA and one NVIDIA Tesla K20x GPU.  Each CPU has 16
AMD Bulldozer cores, each with one floating point unit. All compute nodes are
connected via Cray's Gemini interconnect, which arranges pairs of compute nodes
in a $24\times24\times24$ 3D torus, providing up to 9.6 GB/s
of communication between individual compute nodes. Each pair of compute nodes share a Gemini 
router for the Gemini network, which does not use a centralized switch but
instead passes network packets between neighboring Gemini routers. Blue
Waters is connected to the external Internet via multiple, redundant 40-Gbps
and 100-Gbps connections, with each compute node capable of accessing the
public Internet. The diskless compute nodes and the login nodes share three
distinct, cluster-wide Lustre file systems supplying 26.4 PB of online storage,
with a maximum aggregated transfer speed of more than 1.1 TB/s. 25 dedicated
import/export nodes accept data transfer requests via Globus
Online~\cite{foster:2011} and allow for data to be staged without using resources
on the login or compute nodes. Blue Waters has a total of 
1.63 PB of global memory~\cite{bluewaters:web} or approximately 4 GB per
compute node core, without any swap partition. This supports data-heavy
workloads that benefit from having a large amount of memory, fast, shared file
systems to process data.

Access to  Blue Waters is through three external login nodes that use SUSE Linux
Enterprise Server 11 and require token-based two-factor
authentication. The compute nodes normally provide a light-weight
version of Linux, called Cray Linux Environment (CLE).
This is the mode typically used by high-performance computing applications
that rely on the Message Passing Interface (MPI) for communications among the
compute nodes that make up the job. To simplify transition from a traditional
white box clusters, jobs can also be submitted using the ``cluster compatibility
mode,'' which provides each node with a full-featured Linux environment.
Jobs are submitted to Blue Waters using the Torque/Moab scheduler, which
provides each compute job with a logically consolidated subset of the full $24\times24\times24$
torus in which to run, thus reducing the communication distance between nodes within a job
and reducing network traffic interference between all compute jobs.
Job sizes range from one to more than 22,000 nodes, with typical jobs using two to four-thousand nodes, and with a maximum allowed runtime of
48 hours, after which the scheduler will force termination. 

\begin{figure}[tb]
   \centering
   \includegraphics[width=\linewidth,  clip]{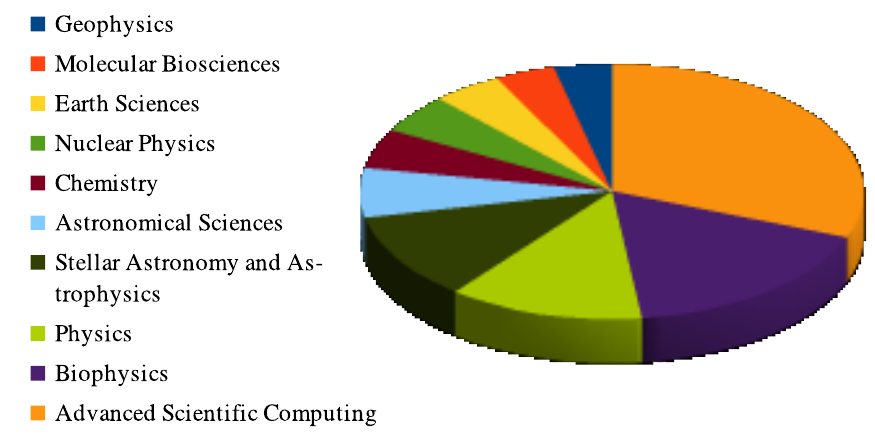}
  \caption{Top ten science areas on Blue Waters in 2017. A total of 291 M node
  hours was allocated. Physics and astrophysics make up approximately 25\% of the
  allocated node hours.
   \label{fig:usage}}
\end{figure}

Figure~\ref{fig:usage} shows the breakdown of the allocated node hours on Blue Waters
by field of science, illustrating the versatility of the cluster~\cite{Bode2013}. Current applications
range from data heavy image processing in the ArcticDEM project~\cite{arcticDEM:web}
to large biomolecular simulations~\cite{zhao2013mature} to numerical relativity.
The project reported in this manuscript adds gravitational wave data analysis to the 
portfolio of science that can use Blue Waters.

\subsection{Open Science Grid}
\label{osg}
The Open Science Grid (OSG), which provides federated compute resources for
data-intensive research~\cite{pordes2007open}, was initially designed primarily to serve US researchers using the
Large Hadron Collider (LHC) at CERN, and is now serving a variety of science
areas.

OSG targets typical high-throughput workloads consisting of spatially small (a few cores to at most one compute node), loosely coupled science jobs that are executed on any of the participating resource providing
clusters~\cite{livny1997mechanisms}.
Since the compute resources are locally owned, rather than being owned by OSG, this brings with it a diverse set of
local policies and priorities that OSG has to support in order to execute
jobs. We use this flexibility to target HTC workloads at Blue Waters, which is
a poster-child HPC resource, favoring large jobs that span many compute nodes.

To target different cluster environments, OSG uses pilot
jobs~\cite{Luckow:2012:TCM:2287076.2287094}  to reserve nodes on the providing
clusters~\cite{sfiligoi2009pilot}. The pilot jobs in turn connect to a
controlling server on the OSG system and request work to be assigned to them.
OSG handles data transfer from the site holding the required data to
the resource providing clusters, so that an application can access the
required files on its own local file system.
Thus OSG provides a virtual, large batch cluster for science workloads by
using pilot jobs. 

Since jobs submitted to the OSG are executed on physical clusters that may have a different computing environment from the submit node, it is important to ensure that the jobs consist of generic binaries and data that can be either carried with the job or staged on demand. Additional requirements to run jobs on OSG include: (i) software should be single threaded, requiring less than 2~GB of memory in each invocation and can run for up to 12 hours. Computations requiring MPI communication will not work on OSG since the infrastructure is distributed; (ii) jobs may be killed and re-started in another site if jobs with higher priority enter the system; (iii) binaries should ideally be statically linked. Languages such as Python and Perl can be used as long as  no special module requirements are needed; (iv) input and output data for each job is limited to 10~GB; and (v) computations requiring a shared file system or complex software deployments are not good matches for OSG. 

The OSG is ideally suited to tackle scientific problems that can be solved by breaking them into a very large number of individual jobs that can run independently, which meets the description of LIGO's most computationally expensive gravitational wave data analysis workflows. OSG is now being used as a universal adapter that allows LSC data analysts to submit their search pipelines using a familiar Condor interface at an LDG site, and then seamlessly run these jobs on external resources. The infrastructure behind this idea is the following: 

\begin{itemize}
\item Users submit jobs using the HTCondor Schedd process at an LDG site.  (HTCondor is a specialized workload management system for compute-intensive jobs, which provides a job queuing mechanism, scheduling policy, resource monitoring, resource management and  priority scheme.   Schedd is an HTCondor system used to submit and queue jobs.)
\item The Glidein Workload Management System (GlideinWMS) Frontend polls the local HTCondor pool to match the required number of user jobs with glideins or worker nodes.  (Glidein is a mechanism used by remote machines to temporarily join a local HTCondor pool.) 
\item Glideins are Condor-based pilots. A pilot system is an infrastructure that creates a virtual private batch system~\cite{Luckow:2012:TCM:2287076.2287094}. Pilots are containers, submitted to the grid, rather than individual user jobs. Once pilots land on a grid resource, they detect local resources. Thereafter, they often fetch, start and monitor user jobs, though jobs can also be sent to them for them to run.   
\item OSG facilities receive the glidein jobs, which now show up as a resource in the HTCondor pool.
\item Jobs are run on OSG worker nodes until completion. If new jobs are scheduled in the HTCondor pool, the process continues until all the jobs of the workflow started at the HTCondor cluster runs to completion.
\end{itemize}

This approach has significantly increased the utilization of additional dedicated, shared, and opportunistic HPC and HTC resources beyond the LDG. These new resources have been heavily exploited in the detection of gravitational waves by aLIGO.

\subsection{LIGO needs}
\label{ligo_needs}

During LIGO's first gravitational wave detection campaign, known as O1, two runtime software environments were used. The LDG (dedicated LIGO Lab and LSC HTC clusters) contributed about 83\% of computational resources. LIGO also harnessed 17\% of O1 computing by using OSG as a universal adapter to external resources (campus/regional shared clusters and NSF-funded supercomputers such as XSEDE and opportunistic cycles from DOE labs and HEP clusters), and to provide elasticity to LIGO computing resources to meet peak or unexpected demand. 

Based on the expected improved sensitivity of aLIGO detectors, and the increased lifetime of detection campaigns, it is expected that
\begin{itemize}
\item O2 aLIGO data analysis will consume more than two times the computational resources than O1 (with two LIGO detectors and a much longer campaign than O1)
\item O3 will require five times as much computing as O1 (2 LIGO detectors and the European gravitational wave detector Virgo, plus a longer detection campaign than O2). Estimates suggest that 0.5 billion SUs\footnote{1SU= 1 aLIGO SU= 1 Intel Xeon E5-2670 2.6GHz CPU core-hour. For reference, 1 aLIGO SU = 0.96 XSEDE SU.} will be consumed during O3. 
\end{itemize}

Given current demand estimates, aLIGO may have sufficient resources throughout 2017-2018 for high priority computing activities. This takes into account more than ninety prioritized gravitational wave searches and detector characterization analyses, and more than sixty pipelines that will be used for these studies. Additional non-LDG resources will be of great benefit to meet peak or unexpected demand, and to unlock new science that may be categorized as high risk-high reward. For instance, leadership facilities such as XSEDE contributed about a third of OSG cycles during O1. Other centers such as the Holland Computing Center at the University of Nebraska stored about 5~TB of input data, and  more than 1~PB of total data volume distributed to jobs during O1.

LIGO currently uses the Pegasus Workflow Management System~\cite{deelman2005pegasus} as a layer on top of DAGMan to manage dependencies. DAGMan (Directed Acyclic Graph Manager) is provided by HTCondor to enforce dependencies between jobs in large workflows, and reliably restart workflows from point of failure.  
Furthermore, many LIGO pipelines use the Grid LSC User Environment (GLUE) LSCSoft~{LAL} package to automate the construction of DAGMan and Pegasus source files. These tools will enable the use of non-LDG resources going forward. 

PyCBC~\cite{2016CQGra..33u5004U} and GstLAL~\cite{2014PhRvD..89b4003P} are the two gravitational wave search pipelines that have been the largest consumers of computing resources since O1. PyCBC is the most computationally intensive pipeline, and the only production pipeline that currently runs on OSG. Other major pipelines (continuous wave burst (cWB)~\cite{Sergey:2016}, LALInference~\cite{bambiann:2015PhRvD} and Bayeswave~\cite{corn:2015CQGra}) will be able to run on OSG resources in the near future. The following subsection describes in  detail the PyCBC pipeline.

\subsection{PyCBC}
\label{pycbc}

PyCBC is a Python software package that is used to perform matched-filtering,  coincident searches of gravitational wave signals in LIGO and Virgo data~\cite{2016CQGra..33u5004U}. PyCBC is one of the most computationally demanding gravitational wave 
search workflows used by the LSC~\cite{2016CQGra..33u5004U}.

PyCBC has been used in off-line mode 
for the validation of the first two gravitational wave transients reported  
by the LSC~\cite{DI:2016,secondec}. A new, low-latency 
version of this package---PyCBC Live---was used to
carry out the detection of the third gravitational wave event that was recently  reported by the LSC~\cite{thirddetection}. We use this 
workflow as a case study, since it has a mature workflow
planner---implemented in Pegasus~\cite{deelman2005pegasus}---and Python bundling
that enables its use on LDG, OSG and XSEDE resources~\cite{2017Weitzel}.

To analyze one day of LIGO data, a PyCBC workflow typically requires about one hundred thousand jobs. Each of these jobs will read one or two frame files that contain the calibrated output of the LIGO detectors. These files are about 400~MB in size, and contain 4096 seconds of LIGO data. PyCBC workflows have tasks that typically run on single cores, and with execution times that range between a few and tens of hours. For the workflows used in this work, we have found that data transfer per task takes, on average, about 5.5 seconds. On the other hand, each task has an average compute time of three to five hours. Therefore, this analysis is CPU-limited during the actual computation phase.
Nevertheless, each job requires several hundred MB of data that it reads from disk
multiple times. Fast Fourier Transform (FFT) computations of the matched-filtering algorithm, and signal consistency tests dominate the operation counts, which increase with the size of template banks---currently including in excess of  \(\sim 3\times10^5\) modeled waveform---and the amount of input data.

To run PyCBC on OSG, we configured an LDG center that is readily accessible to the authors of this article. This work is described in the following section.

\subsection{Configuration of an LDG Tier-1 OSG center}
\label{ldas-osg}
The heterogeneous PyCBC workflow requires an instance of Schedd that can submit jobs to a regular LDG-style pool with a shared filesystem for pre- and post-processing jobs, and to any 
other combination of resources---opportunistic computing
resources, NSF supercomputers and commercial clouds---through GlideinWMS.

The shorter and I/O-intensive pre- and post-processing PyCBC jobs (as well as those with LDG software or service dependencies) are regular, vanilla, universe Condor jobs that are submitted to a `local' LDG HTC cluster (that is not part of OSG), and are defined as non-glide-in jobs. On the other hand, the most CPU-intensive and portable portion of the workflow, the ``pycbc\_inspiral jobs'', are submitted to the OSG pool. 

\begin{figure}[tb]
   \centering
   \includegraphics[width=\linewidth,  clip]{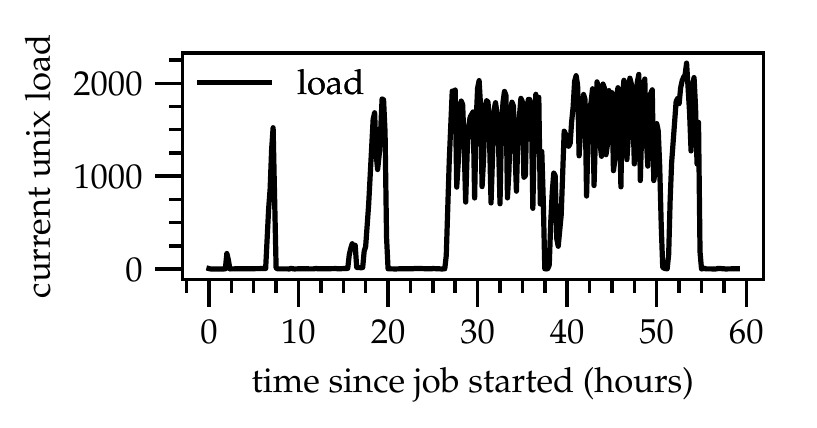}
  \caption{The plot shows the number of tasks that run at the LDG OSG submission machine for different PyCBC workflows. The first blip to the left represents the small PyCBC workflow that we use for validation purposes, and which is used by LDG clusters to perform GitHub Travis CI tests for any software updates to PyCBC. All the workflows shown in this plot ran to completion using only OSG resources.
   \label{fig:ganglia_plot}}
\end{figure}

In order to run PyCBC workflows on OSG resources, we configured an LDG Tier-1 Center, located at the California Institute of Technology, with OSG submission capabilities. While a LDG Tier-3 Center had previously been configured to submit OSG jobs, which enabled users of that center to submit jobs to OSG, this was the first time this had been done for a Tier-1 Center, which enables all LDG users to use Blue Waters for LSC analyses. After successfully running several small PyCBC workflows, we ramped up the size of the workflows from several hundred to several tens of thousands of tasks. We found that for the largest workflows, hundreds of intermediate Pegasus files are stored under the \texttt{/tmp} directory on the worker nodes at the local LDG cluster. These files in \texttt{/tmp} fill up the root directory, and cause the Condor starters (which handle all the details of starting and managing condor jobs) to intermittently fail, which leads to the eventual failure of the workflow. To address this porblem we changed the location where the intermediate Pegasus files are stored in the worker node from \texttt{/tmp} to \texttt{/local/\$USER}. To accomplish this, we pass the following command line arguments to the PyCBC command \texttt{pycbc\_submit\_dax}:

\vspace{2mm}
\small{
\texttt{\$ pycbc\_submit\_dax$\backslash$ }

\texttt{--accounting-group GROUP$\backslash$ }

\texttt{--dax \${WORKFLOW\_NAME}.dax$\backslash$}

\texttt{--execution-sites osg$\backslash$ }

\texttt{\textbf{--append-site-profile} "local:env|PEGASUS\_WN\_TMP:/local/\${USER}"$\backslash$} 

\texttt{\textbf{--append-site-profile} "local:env|TMPDIR:/local/\${USER}"$\backslash$}

\texttt{\textbf{--append-site-profile} "local:env|TMP:/local/\${USER}"$\backslash$}

\texttt{--append-pegasus-property 'pegasus.transfer.bypass.input.staging=true'$\backslash$}

\texttt{--remote-staging-server `hostname -f`}
}
\vspace{2mm}

where \texttt{PEGASUS\_WN\_TMP:/local/\${USER}} indicates that the numerous intermediate Pegasus files created during the execution of the workflow should be stored in \texttt{/local/\${USER}}. Before adding these environment variables, we were only able to run PyCBC workflows on OSG with the dataset used by LDG clusters to perform GitHub Travis CI tests for any software updates to PyCBC. This dataset spans 0.083 days worth of LIGO data that covers GW150914 data, which represents the first gravitational wave transient detected by aLIGO, see the first blip to the left of Figure~\ref{fig:ganglia_plot}. With the aforementioned modifications, we were able to run PyCBC workflows that are more than 40 times larger, see Figure~\ref{fig:ganglia_plot}. For the largest workflows, several thousands of coincident jobs were run concurrently using OSG resources.

Having configured an LDG Tier-1 Center with an OSG submission machine to which NCSA LIGO scientists have access to, we proceeded to run a PyCBC workflow on Blue Waters via OSG. To accomplish this task we used containers. Our team is one of the early adopters of containers to run scientific workflows on Blue Waters, and the first team that has successfully used both OSG and containers to run scientific workflows on Blue Waters. Consequently, this work also represents the first time Blue Waters, containers, OSG and an LDG Tier-1 cluster have been used to successfully run LIGO gravitational wave searches.

\subsection{Containers in the Blue Waters Supercomputer}
\label{shifter}

PyCBC has been thoroughly tested by LIGO on a number of different 
clusters. However, all of them them used a variant of the \texttt{RedHat} or \texttt{Debian} operating
systems on their compute nodes, where specific versions of these operating systems were used to certify the software stack. 
This presented a challenge, as Blue Waters does not operate on these tested Linux variants, but on a lightweight Linux variant based on SUSE.

To overcome this challenge, Blue Waters adopted Shifter~\cite{Jacobsen2015ContainTU}
as its container solution. Shifter accepts Docker~\cite{docker:web} image files,
and converts them into a disk image suitable for concurrent use by multiple compute nodes
of Blue Waters. Shifter ensures that the system-wide, parallel file systems are visible inside 
of the container; that MPI can be used; and that Blue Waters' security policy is enforced on the
container.

Using Shifter, each job 
runs inside a previously developed and tested Docker container. 
For the work in this paper, we built a Docker container with a CentOS6 image, in which PyCBC has 
been shown to work at other compute sites.

\begin{figure}[tb]
   \centering
   \includegraphics[width=\linewidth,  clip]{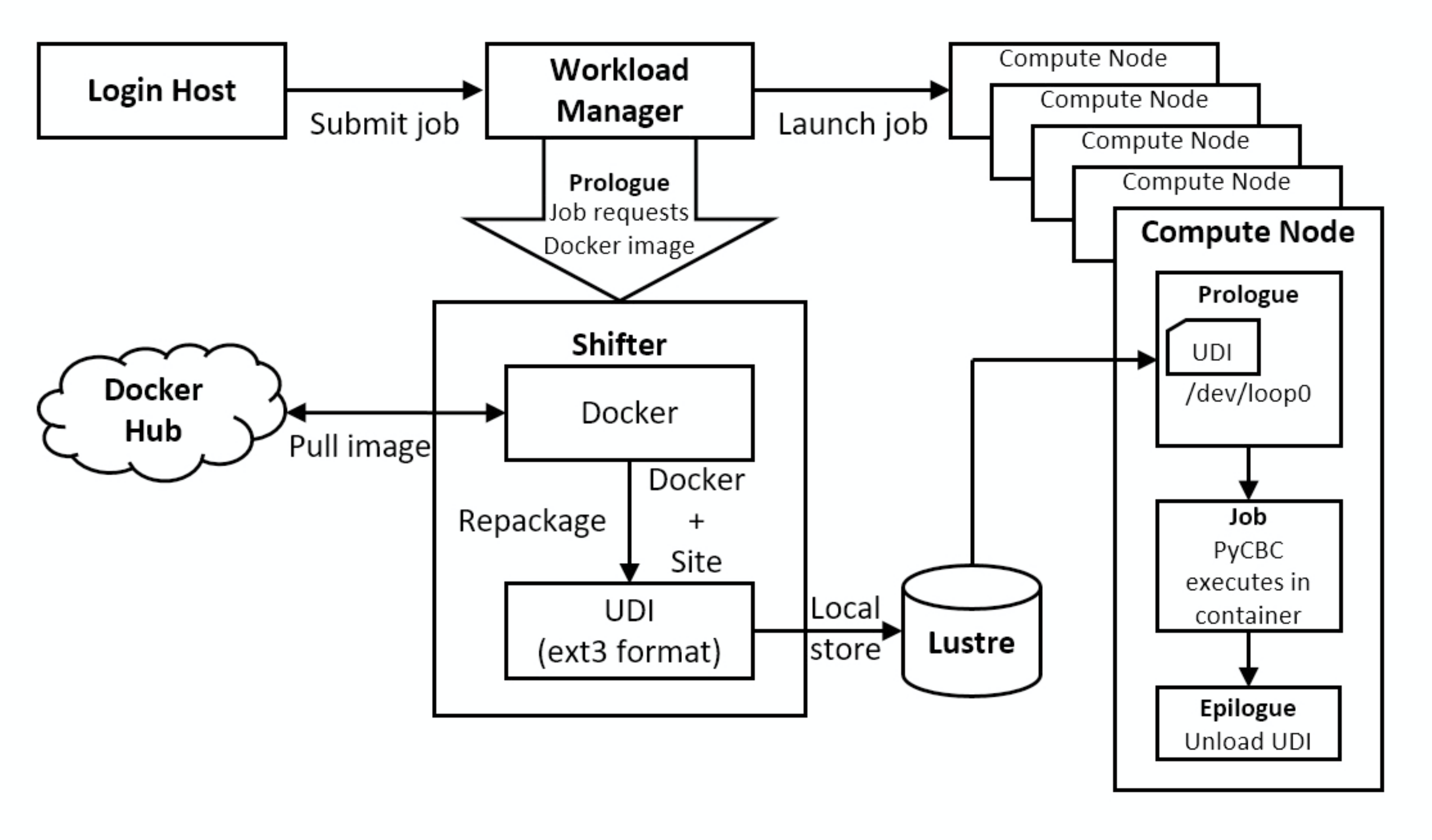}
  \caption{Use of Shifter to run LIGO workflows on Blue Waters. 
   \label{fig:shifter_plot}}
\end{figure}

Shifter allows multiple compute
nodes to share this disk image, rendering file accesses to files in the container into accesses to the single disk image on Blue Waters. 
On a large system-wide Lustre
installation as is used on Blue Waters \cite{Kramer2015}, this significantly improves performance
since opening or closing a file is handled by a single, cluster-wide metadata server that has to 
handle all requests from all processes in the cluster. Since Shifter uses a disk image
to encapsulate all files of the container, file open and close events inside of the container are 
not passed to the system wide metadata server but are bulk data accesses that are handled by the much more numerous 
object data servers of Blue Waters' Lustre file system. This is shown in Figure~\ref{fig:shifter_plot}.

To run PyCBC workflow on Blue Waters via OSG, we need the following
modifications to the PyCBC command \texttt{pycbc\_submit\_dax}:

\vspace{2mm}
\small{
\texttt{\$ pycbc\_submit\_dax$\backslash$ }

\texttt{--accounting-group GROUP$\backslash$ }

\texttt{--dax \${WORKFLOW\_NAME}.dax$\backslash$}

\texttt{--execution-sites osg$\backslash$ }

\texttt{\textbf{--append-site-profile} "local:env|PEGASUS\_WN\_TMP:/local/\${USER}"$\backslash$} 

\texttt{\textbf{--append-site-profile} "local:env|TMPDIR:/local/\${USER}"$\backslash$}

\texttt{\textbf{--append-site-profile} "local:env|TMP:/local/\${USER}"$\backslash$}

\texttt{--append-pegasus-property `pegasus.transfer.bypass.input.staging=true'$\backslash$}

\texttt{--append-site-profile `osg:condor|Requirements:(IS\_GLIDEIN=?=True)'$\backslash$}

\texttt{--append-site-profile `osg:condor|+DESIRED\_SITES:"BlueWaters"'$\backslash$}

\texttt{--remote-staging-server `hostname -f`}
}
\vspace{2mm}

With these modifications, pre-- and post- processing PyCBC jobs will be directed to the local LDG Condor pool, whereas ``\texttt{pycbc\_inspiral\_jobs}" will be  routed to Blue Waters. To do this latter part, pilot jobs are submitted from within Blue Waters using

\vspace{2mm}

\small{

\texttt{\#!/bin/bash}

\texttt{\#PBS -N bluewaters.job}

\texttt{\#PBS -v UDI=\$USER/centos6:osg-wn-client-v1}

\texttt{\#PBS -l nodes=1:ppn=1}

\texttt{\#PBS -l gres=ccm\%shifter \#\#PBS -l walltime}

\texttt{module load shifter}

\texttt{mount | grep /var/udi}

\texttt{export CRAY\_ROOTFS=UDI}

\texttt{cd \$PBS\_O\_WORKDIR}

\texttt{mkdir -p /dir/\$USER/\$PBS\_JOBID}

\texttt{export SCRATCH=/dir/\$USER/\$PBS\_JOBID}

\texttt{aprun -n 1 -N 1 glidein\_startup.sh $\backslash$ }

\texttt{<input.data> output-shifter.\$PBS\_JOBID 2>$\backslash$}

\texttt{outerr-shifter}

\texttt{\$PBS\_JOBID}

}

\vspace{2mm}

Using this approach, we succeeded in running LIGO workflows that are submitted from an LDG cluster, but run on Shifter using Blue Waters computing resources, and where OSG plays the role of a universal converter between the LDG and Blue Waters.

\section{Validation}
\label{check}

To validate our results, we first ran a small PyCBC workflow on OSG facilities using the dataset utilized by GitHub Travis CI tests on LDG clusters. This dataset and the results obtained from this analysis have been thoroughly cross-checked using LDG and OSG resources. Therefore, we used the results of this analysis as a validation workflow.

Having a baseline for comparison, we run a PyCBC workflow on Blue Waters using the same validation dataset, and thoroughly checked that the results reported in both independent analyses were identical. We found that the sixteen parameters describing the identified gravitational waves that are reported by the workflow for the top twenty loudest events were identical. 

Thereafter, we repeated the same exercise running ten times larger PyCBC workflows both on OSG and Blue Waters, and confirmed that the results were consistent. Upon confirming that our computational infrastructure works in a stable manner, and that we are able to accurately reproduce results obtained with OSG resources, we stress-tested this new framework with a production run workflow, as we describe in the following section.

\section{Scalability}
\label{scale}

\begin{figure}[tb]
   \centering
   \includegraphics{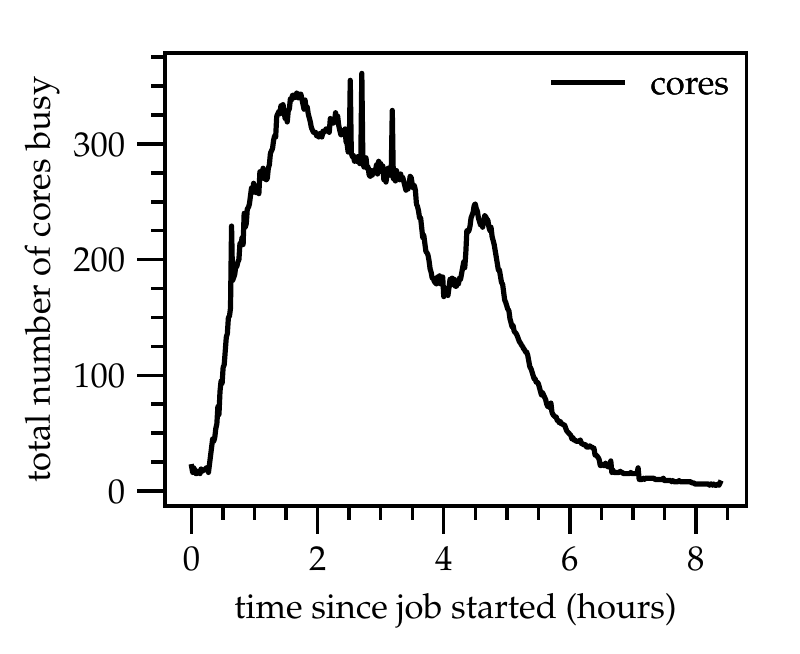}
  \caption{Utilization of the 400 cores in the 25 Blue Waters nodes reserved by the LIGO pilots for a small scale test.
   We started 5 sets of pilot jobs with 5 nodes (80 cores) in each to supply
   resources to OSG. 3 of the pilots started at the same time with the other
   two delayed by approximately 20 and 50 minutes respectively. It takes
   almost 2 hours before all PyCBC jobs are started and the core utilization
   never reaches 100\% of the 400 cores provided by the pilots. Pilot jobs
   are released once there are no more PyCBC jobs waiting to be started with
   only a single pilot still active after 5 hrs.
   \label{fig:ovis-load}}
\end{figure}

Figure~\ref{fig:ovis-load} shows the number of cores that are busy as a
function of time for a small workload utilizing 25 compute nodes on Blue
Waters. The 25 nodes are distributed among 5 independent pilot jobs, adding
granularity to the system since pilot jobs terminate once there are no longer
any PyCB jobs waiting to be scheduled. On the other hand, startup time is
delayed since not all pilot jobs start at the same time resulting in slow
rise of the number of utilized cores. The pilot jobs 
include both the movement of gravitational wave data to Blue Waters, 
which is followed by the computational intensive ``pycbc\_inspiral" tasks.

\begin{figure}[tb]
   \centering
   \includegraphics{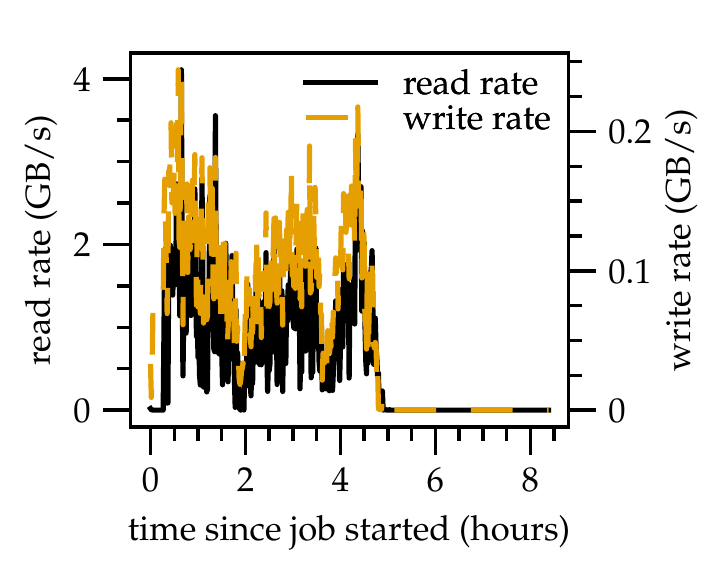}
   \caption{Aggregated I/O rate during the workflow. IO rate is highest while
    the pilots start up until all PyCBC jobs enter the computational phase at
    2 hrs after the pilots started. At the 5 hrs mark IO rate drops as the no
    more data is read by newly started jobs.
   \label{fig:ovis-iorate}}
\end{figure}

Figure~\ref{fig:ovis-iorate} shows the aggregated read and write speed during
the workflow. The read rate is much higher than the write rate since each
PyCBC job reads its input multiple times. Since each PyCBC is independent of
the others, these rates scale linearly with the number of cores used.

\begin{figure}[tb]
   \centering
   \includegraphics{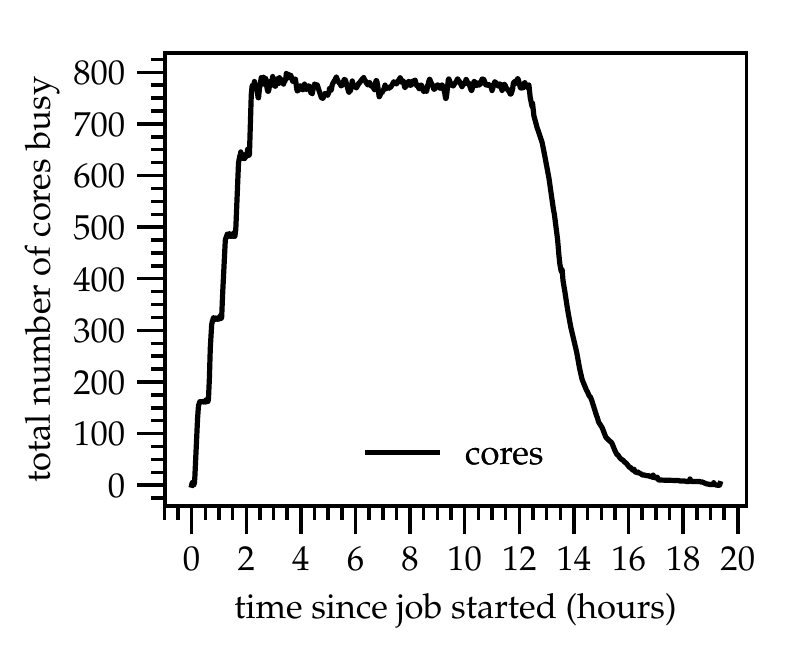}
  \caption{Utilization of the 800 cores in the 50 Blue Waters nodes reserved by the LIGO pilots for our large scale test.
   To avoid overloading the OSG server, we start pilots that each use 10 nodes, with a 30-minute delay between each pilot launch.
   This can be clearly seen in the plateaus in core usage as we ramp up the number of pilots.
   Steady state is reached after about 2.5 hours and lasts until approximately 12 hours after the first pilot was started.
   After that, the pilots drain as the jobs slowly finish until the last job completes at about 19 hours after pilot start time.
   \label{fig:load}}
\end{figure}

We found that scaling PyCBC workflows to tens of thousands of
concurrent jobs poses a serious challenge
to the underlying OSG infrastructure because all the jobs  start at possibly
the same time.
Throttling the initial staging of data from the LDG to Blue Waters,
for example, becomes mandatory even when using a modest number of nodes
(in the hundreds), as
otherwise the LDG file server is overwhelmed by transfer requests.
This can be seen in Figure~\ref{fig:load}, which shows the number of cores that are busy as a function
of time for a collection of pilots running on 50 nodes. Here we started
pilot jobs in groups of 10 nodes with a 30-minute delay between the groups. This can
be clearly seen in the 5 plateaus until the usage reaches
its peak approximately 2.5 hours after the pilots were started. A detailed inspection of this graph shows that staging the data for 10 worker nodes took
almost 15 minutes. Thus, staging data is in fact a bottleneck for our
current implementation. We expect to improve on this in future work by switching
to third party GridFTP transfers using Blue Waters' dedicated I/O nodes rather than
the generic compute nodes. This approach is expected to reduce staging time by at least an order of magnitude. 

Similarly, the Cray Linux Environment (CLE) runtime environment used on Blue Waters is heavily tuned
towards typical HPC applications, and differs from a
commodity hardware cluster software infrastructure found in HTC clusters or smaller HPC centers. Naively using the implementation of the procedures for running PyCBC workflows on XSEDE systems initially caused the performance of the workload per node to be 1/16 of its
capability, because process to core binding is the default on Blue Waters, but not on commodity clusters. This caused all 16 processes on each node to compete for a single core on the node, rather than each running on a separate core. Thus, instead of using

\vspace{2mm}

\texttt{aprun -n 1 -N 1 glidein\_startup.sh}\,,

\vspace{2mm}

\noindent in the script used to start up the pilot jobs on Blue Waters (see the last script of Section~\ref{shifter}), which sets the depth (\texttt{d}) to one (\texttt{-d=1}) and hence binds all jobs to a single core, leading them to compete for resources, Cray systems require the following modification:

\vspace{2mm}

\texttt{aprun -n 1 -N 1 -d 32 glidein\_startup.sh}

\vspace{2mm}

\noindent This ensures that the depth corresponds to the number of physical cores on a node (\texttt{-d=32}), which allows all the cores to be used effectively.

\begin{figure}[tb]
   \centering
   \includegraphics{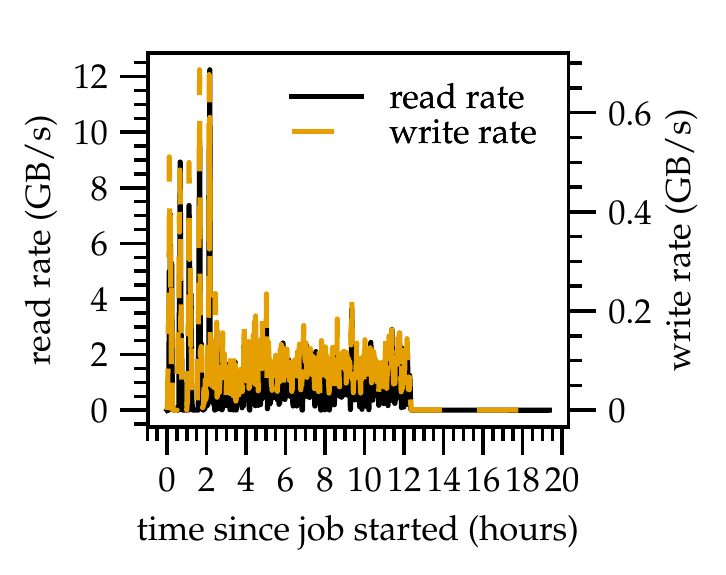}
   \caption{Aggregated I/O rate during the the workflow. The initial spikes
   correspond to pilots starting up and downloading their work package from
   the OSG server. Note the different scaling for read (left axis) and write
   (right axis). The read rate is significantly higher because the same file is read
   multiple times by the workflow. After about 12 hours, no more new jobs are
   started and I/O drops as no additional work packages are downloaded.
   \label{fig:iorate}}
\end{figure}

Particularly during the initial startup of the jobs, the peak I/O rate
is appreciable, and it scales linearly with the number of tasks started.
We started the pilot jobs in different groups so that a large number of pilot jobs
starting at the same time on Blue Waters didn't overwhelm the LDG storage
server supplying analysis data to the PyCBC jobs.
The read data can be seen in the five spikes in Figure~\ref{fig:iorate} in the first 2.5
hours since the job started, which correspond to the time when pilot jobs are
launched. The write
rate, on the other hand, is much lower as each job writes only a small
amount of data to disk once it finishes. This can also be clearly seen in
Figure~\ref{fig:iorate} which shows aggregated I/O rates during the lifetime of the
pilots.

We were able to overcome these challenges and run  PyCBC workflows with several thousand ``pycbcb\_inspiral'' jobs with relative ease. The extension of this framework to effectively handle workflows that have several tens of thousands of ``pycbcb\_inspiral'' jobs will be reported in an upcoming publication. 

\begin{figure}
\centering
\includegraphics{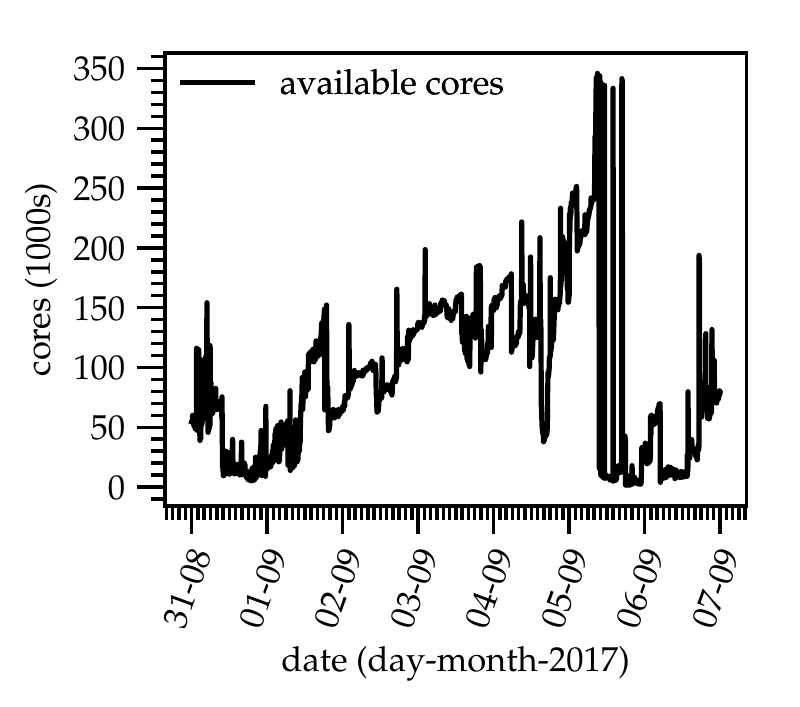}
\caption{Available cores during the week August 31, 2017 to September
7, 2017. A full system job was scheduled on September 5, leading to
nodes draining in anticipation of this job staring as early as
September 1. The total OSG usaable compute resources during this time
are $15\times10^6$ core-hours.}\label{fig:avail}
\end{figure}

Compared to other HPC clusters in the OSG and in particular network, Blue
Waters is by far that largest system in terms of raw core count as shown in
table~\ref{tab:corecounts}
\begin{table}[tb]
\centering
\begin{tabular}{l*{6}c}
\textbf{cluster} &
\textbf{Blue Waters} &
\textbf{Comet} &
\textbf{Atlas} \\
&
NCSA &
SDSC &
Max Planck Society \\
\textbf{cores} &
362240 &
46656 &
31000 \\
\hline\hline\\
\textbf{cluster} &
\textbf{LDG} &
\textbf{NEMO} &
\textbf{UICAA}\\
&
multiple &
University of&
India \\
&
locations &
Milwaukee &
\\
\textbf{cores} &
18780 &
3392 &
2520 \\\hline\hline
\end{tabular}
\caption{Number of compute cores available in clusters in LDG and OSG. LDG
clusters are dedicated to LIGO while OSG clusters provide compute cycles on a
best effort basis, mixing LIGO workloads with regular HPC workloads.}
\end{table}
However since OSG workflows share the clusters with other paralell workflows
the number of available cores to be used by OSG in an opportunistic manner
more realistically shows the amount of computation that can be provided to
LIGO. Figure~\ref{fig:avail} shows that number of avaialble cores on Blue
Waters during the week of August 31, 2017 to September 7, 2017. During this
week a full system job was scheduled on September 5 and the cluster can be
seen to drain (empty nodes) for this large job already on September 1. During
the period shown, $15\times10^6$ core-hours of computer time were usable by
OSG.

\section{BOSS-LDG for gravitational wave discovery campaigns}
\label{application}

\begin{figure}[tb]
   \centering
   \includegraphics{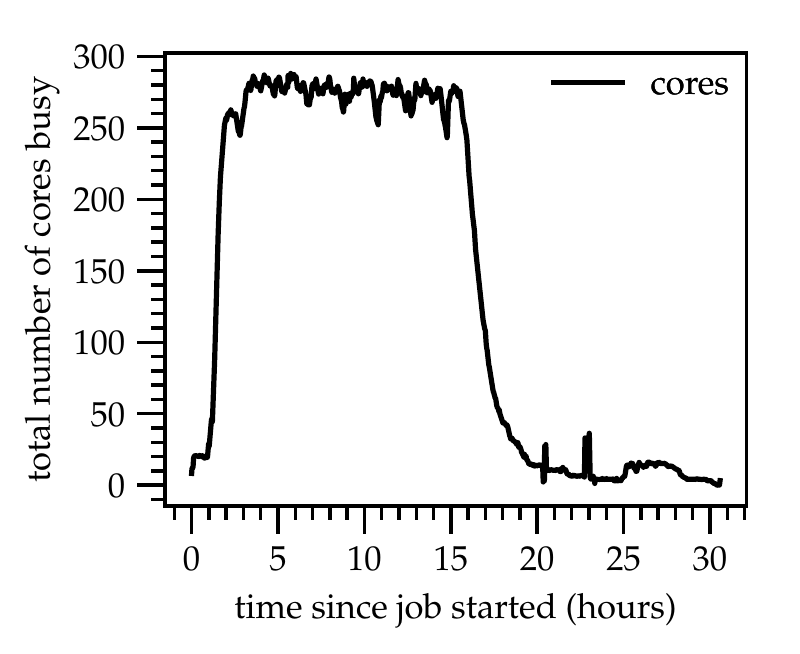}
   \caption{Core utilization of ``pycbc\_inspiral" jobs in a large scale, gravitational wave search run on Blue Waters during the last few weeks of aLIGO's second discovery campaign. 
   \label{fig:pro_run}}
\end{figure}

We started using \texttt{BOSS-LDG} for aLIGO gravitational wave searches
from August 21st 2017, and covered the last several weeks of aLIGO's second discovery
campaign. For these production runs, we used the distributed data access infrastructure
described in~\cite{2017Weitzel}. To determine the right balance between the number of
pilot jobs and core utilization, we submitted pilot jobs that each used 
10, 25, 50 and 100 nodes. We found that an optimal combination consists of a large number of pilot jobs, each using 
10 nodes. Figure~\ref{fig:pro_run} presents a snapshot of core utilization of
 ``pycbc\_inspiral" jobs that were run on Blue Waters 
 during aLIGO's second discovery campaign.

\begin{figure}[tb]
   \centering
   \includegraphics[trim={4cm 6.5cm 3cm 6cm},width=0.35\textwidth]{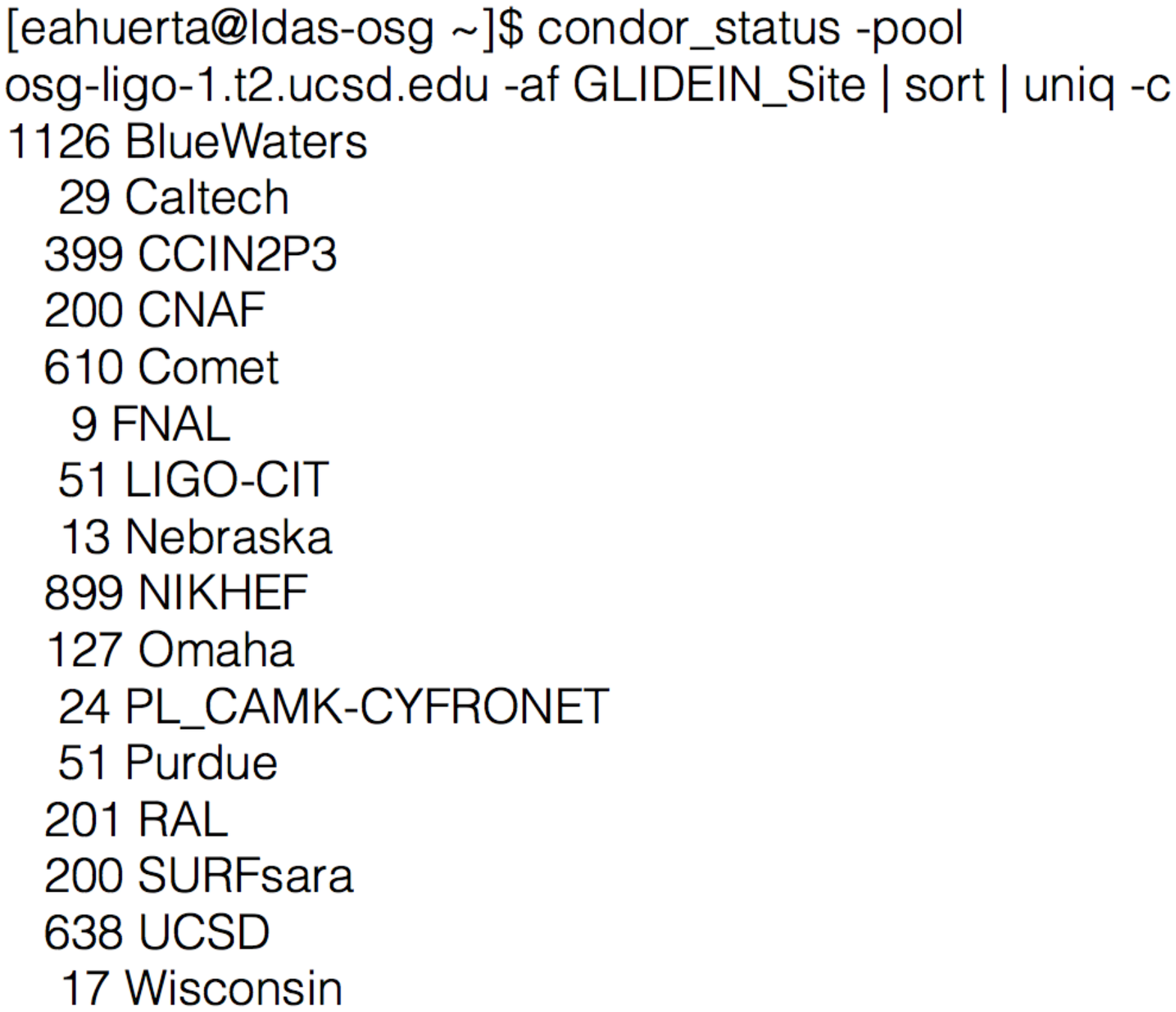}
   \caption{Blue Waters became, at various points during the last few weeks of aLIGO's second discovery campaign, the peak contributor of computational resources to gravitational wave data analysis. 
   \label{fig:peak}}
\end{figure}

As shown in Figure~\ref{fig:peak}, this approach enabled Blue Waters to be the peak 
contributor of computational resources for gravitational wave data analysis at various
points during the last few weeks of aLIGO's second discovery campaign.  

\section{Related Work}
\label{related}

Other projects have aimed at using HPC clusters to handle what is essentially a HTC-type workflow.
Within the LHC, the ATLAS experiment is making use of leadership HPC resources using PaNDA and has
shown this works well to backfill into empty compute nodes on Titan~\cite{Nilsson:2014tna},
with hardware similar to that of Blue Waters, making an estimated 300M core hours per year available to the ATLAS experiment. Additional work is currently under way to use Shifter and OSG on Blue Waters for the LHC, and this began just after our
own project, building on what we have done. Other projects have used workflow
managers without Shifter on Blue Waters, for example the ArcticDEM project~\cite{arcticDEM:web} uses
Swift~\cite{WILDE2011633}, and the Southern California Earthquake Center's CyberShake
project~\cite{cui2013accelerating} uses Pegasus~\cite{deelman2005pegasus}. Finally some of the authors of this paper were involved in 
porting the OSG\slash LIGO software stack to XSEDE HPC resources in the past~\cite{2017Weitzel}.

Our approach differs from these previous uses by combining Shifter with OSG to provide the ability to use an project-certified
software environment on Blue Waters for an HTC workflow, using LIGO as a test case.  This removes the need for porting the application software and much of the environment, which could lead to additional validation activities in some science communities. This also supports an easier path towards future reproducibility.

\section{Conclusion}
\label{end}

We have developed a novel computational infrastructure
to connect the LIGO Data Grid to the Blue Waters Leadership supercomputer. 
To accomplish this work, we configured an LDG Tier-1 
Center to handle heterogeneous
LIGO workflows, and showed that we can run  
PyCBC workflows to completion via OSG. At the other end of the spectrum, 
we increased the flexibility and versatility of the Blue 
Waters supercomputer by enabling the use of Shifter, and then made use of this new capability. 
Thereafter, we configured, for the first time, the Blue Waters
supercomputer as an OSG resource for large-scale gravitational wave data analysis, allowing it to run incoming jobs 
from the LIGO Data Grid. We have tested this novel 
framework, and have shown that we can successfully run one 
of the most computationally demanding gravitational wave search packages via OSG submission mechanisms. 

We have throughly checked the results obtained from running PyCBC 
workflows both on OSG resources and on the Blue 
Waters supercomputer, and found that the results were scientifically identical.  Specifically, the top twenty significant triggers (potential gravitational wave candidates) in both analyses have identical numerical results.

The framework we introduce in this article represents the first time Open Science Grid, Shifter, and Blue Waters are unified to tackle a scientific problem and, in particular, it is the first time a framework of this nature is used in thse context of large scale gravitational wave data analysis. We have already used this framework to search for gravitational wave transients during the last few weeks of aLIGO's second discovery campaign.
This framework
can be readily used to run other scientific workflows on the Blue Waters supercomputer, if they meet the following requirements: they are a good match to the OSG infrastructure (see Section~\ref{osg}), the software can be containerized, and a workflow manager can be used to monitor the workflow from end to end, i.e., Pegasus, Swift, etc. This is a minimal set of requirements that may be easily met by existing OSG users, who may already use portable, self-contained software that could be containerized. 
(Note that we use two-factor authentication when submitting the pilot jobs to Blue Waters, so users who want to use this framework are required to obtain a Blue Waters allocation, which is a reasonable requirement to exploit this HPC facility.)

This computational framework 
is under intense development. In an upcoming publication, we will describe a method to overcome existing challenges related to scaling LIGO workflows to fully exploit the unique capabilities of the Blue Waters supercomputer. This work will enable deeper and faster gravitational wave searches, paving the way to accelerate discovery not only for astronomical observatories such as aLIGO, but also for a rich ecosystem of users interested in leveraging this new framework to carry out large scale data analytics research on Blue Waters combining Shifter and OSG. 

In addition, anticipating a change in how containers are supported, we will also investigate the use of Singularity~\cite{sing:2017} containers---specifically tailored to encapsulate the user space environment to facilitate portability and reproducibility, and which do not allow root escalation or user contextual changes---in our framework.

\section*{Acknowledgements}
\label{acknowledgements}
This research is part of the Blue Waters sustained-petascale computing project, which is supported by the National Science Foundation (awards OCI-0725070 and ACI-1238993) and the State of Illinois. Blue Waters is a joint effort of the University of Illinois at Urbana-Champaign and its National Center for Supercomputing Applications. We thank Brett Bode, Juan Barayoga, Greg Bauer, Brian Bockelman, Mats Rynge and Karan Vahi for fruitful interactions.




\bibliography{references}
\bibliographystyle{IEEEtran}

\end{document}